\documentstyle[12pt,axodraw]{article}

\newcommand{\be}{\begin{equation}}
\newcommand{\bea}{\begin{eqnarray}}
\newcommand{\ba}{\begin{array}}
\newcommand{\bean}{\begin{eqnarray*}}
\newcommand{\ee}{\end{equation}}
\newcommand{\eea}{\end{eqnarray}}
\newcommand{\ea}{\end{array}}
\newcommand{\eean}{\end{eqnarray*}}

\def \dsl {\partial \kern-.55em{/}}
\def \Dsl {D \kern-.65em{/}}
\def \nonSM {SM \kern-.75em{/}} 
\def \qsl {q \kern-.45em{/}}
\def \slp {p \kern-.45em{/}}
\def \ksl {k \kern-.45em{/}}

\def\grA{
\begin{picture}(150,80)(0,0)
\Oval(0,0)(20,70)(0)
\Gluon(0,20)(0,-20){2}{9}
\Photon(-35,17.32)(-65,70){-5}{7}
\Line(-88,6)(-70,6)
\Line(-88,-6)(-70,-6)
\Line(70,6)(88,6)
\Line(70,-6)(88,-6)
\BCirc(-70,0){8}
\BCirc(70,0){8}
\Text(-60,18)[c]{$x_1p_1$}
\Text(-54,-22)[c]{$x_2p_1$}
\Text(54,22)[c]{$y_1p_2$}
\Text(54,-22)[c]{$y_2p_2$}
\Text(-72,72)[c]{$V$}
\Text(-40,50)[c]{$q$}
\Text(-95,0)[c]{$p_1$}
\Text(95,0)[c]{$p_2$}
\end{picture}
}

\def\grC{
\begin{picture}(150,80)(0,0)
\Oval(0,0)(20,70)(0)
\Gluon(0,20)(0,-20){2}{9}
\Photon(-35,-17.32)(-65,-70){-5}{7}
\Line(-88,6)(-70,6)
\Line(-88,-6)(-70,-6)
\Line(70,6)(88,6)
\Line(70,-6)(88,-6)
\BCirc(-70,0){8}
\BCirc(70,0){8}
\end{picture}
}
 
\def\grB{
\begin{picture}(200,80)(0,0)
\Oval(0,0)(20,70)(0)
\Gluon(0,20)(0,-20){2}{9}
\Photon(35,17.32)(5,70){-5}{7}
\Line(-88,6)(-70,6)
\Line(-88,-6)(-70,-6)
\Line(70,6)(88,6)
\Line(70,-6)(88,-6)
\BCirc(-70,0){8}
\BCirc(70,0){8}
\end{picture}
}

 \def\grD{
\begin{picture}(200,80)(0,0)
\Oval(0,0)(20,70)(0)
\Gluon(0,20)(0,-20){2}{9}
\Photon(35,-17.32)(5,-70){-5}{7}
\Line(-88,6)(-70,6)
\Line(-88,-6)(-70,-6)
\Line(70,6)(88,6)
\Line(70,-6)(88,-6)
\BCirc(-70,0){8}
\BCirc(70,0){8}
\end{picture}
}

\def\pic{\bea
\hspace*{4cm}\grA & \hspace*{2cm}\grB & \nonumber \\
\hspace*{4cm}\grC & \hspace*{2cm} \grD & \nonumber
\eea
}

\catcode`@=11
\newcount\@tempcntc
\def\@citex[#1]#2{\if@filesw\immediate\write\@auxout{\string\citation{#2}}\fi
  \@tempcnta\z@\@tempcntb\m@ne\def\@citea{}\@cite{\@for\@citeb:=#2\do
    {\@ifundefined
       {b@\@citeb}{\@citeo\@tempcntb\m@ne\@citea\def\@citea{,}{\bf
?}\@warning
       {Citation `\@citeb' on page \thepage \space undefined}}%
    {\setbox\z@\hbox{\global\@tempcntc0\csname
b@\@citeb\endcsname\relax}%
     \ifnum\@tempcntc=\z@ \@citeo\@tempcntb\m@ne
       \@citea\def\@citea{,}\hbox{\csname b@\@citeb\endcsname}%
     \else
      \advance\@tempcntb\@ne
      \ifnum\@tempcntb=\@tempcntc
      \else\advance\@tempcntb\m@ne\@citeo
      \@tempcnta\@tempcntc\@tempcntb\@tempcntc\fi\fi}}\@citeo}{#1}}
\def\@citeo{\ifnum\@tempcnta>\@tempcntb\else\@citea\def\@citea{,}%
  \ifnum\@tempcnta=\@tempcntb\the\@tempcnta\else
   {\advance\@tempcnta\@ne\ifnum\@tempcnta=\@tempcntb \else
\def\@citea{--}\fi
    \advance\@tempcnta\m@ne\the\@tempcnta\@citea\the\@tempcntb}\fi\fi}
\catcode`@=12
 
\begin{document}

\begin{titlepage}

\begin{flushright}
CERN-TH/98-220\\
DTP/98/44 \\
%hep-ph/9807xxx\\
July 1998
\end{flushright}
 
\begin{center}
{\Large {\bf A set of sum rules for anomalous gauge boson couplings}}
\\[2.4cm]
{\large Joannis Papavassiliou}$^a$ 
                           {\large and Kostas Philippides}$^b$\\[0.4cm]
$^a${\em Theory Division, CERN, CH-1211 Geneva 23, Switzerland}\\[0.3cm]
$^b${\em Department of Physics, University of Durham, Durham, DH1 3LE, 
U.K.}
\end{center}
 
\vskip0.7cm     \centerline{\bf   ABSTRACT}  \noindent

The dependence of the differential cross-section for on-shell $W$-pair
production on the  anomalous trilinear gauge couplings invariant under
C and P   is examined.  It is  shown  that  the contributions of   the
anomalous magnetic moments of the $W$ boson due to  the photon and the
$Z$ can  be individually projected out  by means  of two appropriately
constructed polynomials.  The remaining  four anomalous couplings  are
shown  to satisfy a  set   of model-independent  sum  rules.  Specific
models  which predict special  relations among the anomalous couplings
are then studied; in  particular, the composite  model of Brodsky  and
Hiller, and the linear and non-linear effective Lagrangian approaches.
The relations predicted by    these  models, when combined  with   the
aforementioned   sum  rules,   give rise    to   definite predictions,
particular to each model.  These predictions  can be used, at least in
principle, in order to exclude or constrain such models.

\vskip0.7cm 

PACS numbers: 14.70.Fm, 13.40.Gp, 11.55.Hx, 12.60.Rc

\vskip 6cm
e-mail: Joannis.Papavassiliou@cern.ch~,~Kostas.Philippides@durham.ac.uk

\end{titlepage}
%\newpage

\setcounter{equation}{0}
\section{Introduction}

The  possibility  of probing  directly  non-Abelian  vertices  in  the
experiments at the CERN Large  Electron Positron collider LEP2 through
the process  $e^{+}e^{-}\rightarrow W^{+}W^{-}$ \cite{BBL,WG1,WG2}, as
well as the Tevatron \cite{EW} and  the Next Linear Collider (NLC) has
motivated the extensive study of anomalous gauge boson couplings.  The
general  methodology for quantifying the effects  of such couplings on
physical amplitudes  has been presented  some time  ago in the classic
papers by  Gaemers  and Gounaris  \cite{GG} and  Hagiwara  {\sl et al}
\cite{Hagi}.   The  central idea is  to  parametrize  the most general
three  gauge boson  vertex allowed by   Lorentz invariance in terms of
unknown form-factors, compute the  theoretical predictions of relevant
physical  amplitudes  using this vertex,  and  then attempt to extract
information about the structure  of  these form-factors  by  comparing
these theoretical predictions with the experimental data.  In practice
one usually  obtains experimental lower bounds  on  the size of these
form-factors  by   carrying  out a multi-parameter    fit  to the data
\cite{Various}.   This   type of analysis   becomes considerably  more
complicated if  one takes into account  the fact that the two produced
$W$ are  not  stable, but   decay  subsequently through  a variety  of
channels.  The  complexity  of this  problem  necessitates  a detailed
amplitude analysis for the process $e^{+}e^{-}\rightarrow W^{+}W^{-}$;
several  studies based on a variety  of methods, such  as the helicity
amplitude techniques \cite{MPR}, have been carried out, and a plethora
of  complementary  observables  such as   integrated   cross-sections,
angular distributions,  polarized  cross-sections, W density matrices,
and polarization asymmetries, have been proposed \cite{GLMR}.

It would  clearly be useful to  relate directly  some of the anomalous
form-factors to experimentally measurable quantities.  In addition, it
is important to establish a variety of ways for testing experimentally
some of  the characteristic predictions of models  which  give rise to
such anomalous form-factors.  To that  end in this  paper we present a
study complementary to that of \cite{Hagi};  In particular, we compute
the differential  cross-section for the process $e^{+}e^{-}\rightarrow
W^{+}W^{-}$,  under the  assumptions  that:  (i)   the anomalous  form
factors  used  to parametrize  the   non-Abelian vertex  satisfy  {\sl
  separately}  the discrete  symmetries  $C$,  $P$,  and  $T$.   This
assumption, which  is often employed in  the literature, reduces the
number of  possible form-factors from fourteen  down to six.  (ii) the
size of the anomalous couplings is small compared to unity, so that we
may keep only effects linear in them,  and (iii) that  the two $W$ are
strictly on  shell,  i.e.  we  do   not consider the  effect  of their
subsequent decays. Then, following a method developed in \cite{PRW}, a
system of  four independent  algebraic  equations for  the six unknown
anomalous  form-factors is derived.   It turns out  that one can solve
directly two of these equations  and obtain {\sl explicit} expressions
for the two form-factors traditionally  associated with the photon and
$Z$ anomalous magnetic moments of the $W$ in terms of the differential
cross-section. These two expressions are completely model-independent;
therefore they  can  serve as  a testing  ground for  confronting  the
predictions  of different models   for the anomalous  magnetic moments
with   experiment.  The two remaining  equations   constitute a set of
model-independent sum rules \cite{SumRules} for the other four unknown
form-factors.

To demonstrate with specific examples the potential usefulness of such
sum rules  for testing the   viability of models  predicting anomalous
couplings,  or at least   for constraining them,   we first  study the
composite model  proposed by  Brodsky and Hiller  \cite{BH}.  This
model predicts  certain relations between the  anomalous form-factors;
if these relations are fed into the  two remaining equations mentioned
above, the system turns out to be over-constrained, thus leading to two
independent predictions.   These  predictions are particular  to  this
composite model, and could,  at least in principle,  be confronted
with  experiment.   Next we  turn to   an  approach based  on  a gauge
invariant,  effective   Lagrangian  \cite{Schil,Ruj,HISZ,effthe,Goun},
which also provides   relations   among the various   anomalous   form
factors; in particular all form factors are  expressed in terms of the
three   free  parameters   of the  effective   Lagrangian.  When these
relations are combined   with   the two aforementioned  equations   as
before,  one  constraint  emerges,   which  constitutes  a  particular
prediction of this effective Lagrangian approach.

The paper is organized as follows: In Section 1  we briefly review the
general form of the  three-gauge-boson vertex and define the anomalous
form factors which parametrize the deviation of the couplings from the
Standard Model (SM) tree-level values.  In  Section 2 we compute under
the three assumptions mentioned above,  the analytic dependence of the
theoretical  cross-section on these   form  factors, and  show how  to
derive the expressions for the anomalous magnetic  moments, as well as
the  two sum rules.  In  Section  3 we  study some  aspects  of the BH
model, and translate its   predictions at  the level of   experimental
cross-sections.  In Section 4   the effective  Lagrangian  approach is
outlined both for a linear and  non-linear realization of the symmetry
and  its   predictions   are   expressed  in  terms    of experimental
cross-sections.  Finally in Section 5 we summarize our results.

\setcounter{equation}{0}
\section{Anomalous couplings}
In this section we give a brief review of 
the anomalous gauge-boson couplings and establish notation.
The SM three-gauge-boson vertex
$V_{\mu}W^-_{\alpha}W^+_{\beta} $
involving a neutral gauge boson  
$V=\gamma,Z$ 
coupled to a conserved current 
(massless external fermions in the case of the $Z$ boson)
and two on-shell $W$s 
is given by \cite{BGL}
\be
\Gamma_{\mu\alpha\beta}^{V,0}(q,-p_1,-p_2)=
g_V\Gamma_{\mu\alpha\beta}^{0}(q,-p_1,-p_2) ~, 
\ee
with 
 \be
\Gamma_{\mu\alpha\beta}^{0}(q,-p_1,-p_2) = 
(p_2-p_1)_{\mu}g_{\alpha\beta}
+ 2(q_{\beta}g_{\mu\alpha}  -q_{\alpha}g_{\beta\mu})~, 
\label{SMV}
\ee
where 
$g_{\gamma} = g s_w^2$,  $g_{Z} = g c_w^2$, 
$g$ is the $SU(2)_L$ gauge coupling,  and 
$s_w^2=1-c_w^2$ is sine of the weak mixing angle. 
$\Gamma_{\mu\alpha\beta}^0$
satisfies the
following elementary Ward identity:
\be
q^{\mu}\Gamma_{\mu\alpha\beta}^{0}(q,-p_1,-p_2) =
 \Big[p_2^2-p_1^2 \Big]g_{\alpha\beta} ~.
\label{FWI}
\ee
Assuming that the two $W$s are on shell, i.e. $p_1^2=p_2^2=M^2_W$,
we have from Eq.\ (\ref{FWI}) that
\be 
q^{\mu}\Gamma_{\mu\alpha\beta}^{0}(q,-p_1,-p_2)=0 ~.
\label{WIOS}
\ee
The above elementary Ward identities are crucial for the 
gauge invariance of the process 
$f^{+}f^{-}\rightarrow W^{+}W^{-}$
{\sl even} if the external fermionic current is conserved.
To appreciate this fact 
all one has to do is to choose to work
in the axial or planar gauges \cite{axial} instead
of the usual renormalizable ($R_{\xi}$) gauges \cite{FLS}.
In that case,
the bare photon propagator
$\Delta_{0}^{\mu\nu}(q)$ appearing
in graph (1a) assumes the form
\be
\Delta_{0}^{\mu\nu}(q)= \  \Big[g^{\mu\nu}-\frac{\eta^{\mu}q^{\nu}
+\eta^{\nu}q^{\mu} }{\eta q} \Big]\frac{1}{q^2} 
+\eta^2 \frac{q^{\mu}q^{\nu}}{{(\eta q)}^2}\, .
\label{PhotPol}
\ee
The four-vector $\eta_{\mu}$ in the above
expressions is a gauge-fixing parameter;
therefore, physical quantities such as $S$-matrix elements
should be independent of $\eta_{\mu}$.
Evidently, even though all
$\eta$-dependent terms proportional 
to $q^{\mu}$ and $q^{\mu}q^{\nu}$
will vanish when contracted
with the conserved current, the $\eta$-dependent term
proportional to $q^{\nu}$ can only cancel
if the Ward identity of Eq.\ (\ref{WIOS}) holds.

The most general parametrization of the trilinear gauge
vertex consistent with Lorentz invariance  
is given 
in terms of 14 form factors $f_i^V$  \cite{GG,Hagi}:
\bea
\Gamma^V_{\mu\alpha\beta}(q,-p_1,-p_2) &= &
  (1+f_1^V)(p_2-p_1)_{\mu}g_{\alpha\beta}
  + 2 (1+f_3^V)(q_{\beta}g_{\mu\alpha}-q_{\alpha}g_{\beta\mu}) \nonumber
\\
&& -f_2^Vq_{\alpha}q_{\beta}(p_2-p_1)_{\mu}/2M_W^2
-if_4^V (q_{\beta}g_{\mu\alpha}+q_{\alpha}g_{\beta\mu})
 +if_5^V \epsilon^{\mu\alpha\beta\rho}(p_2-p_1)_{\rho} \nonumber \\
 &&+f_6^V\epsilon^{\mu\alpha\beta\rho}q_{\rho}
 +f_7^V (p_2-p_1)_{\mu}\epsilon^{\alpha\beta\rho\sigma}
 q_{\rho}(p_2-p_1)_{\sigma}/M_W^2 ~.
\label{BSMV}
\eea

The first three form factors $f_i$, $i=1,2,3$, preserve
$C$ and $P$ separately. $f_5$ respects $CP$ but violates both $C$ and
$P$. 
The rest of the form factors violate $CP$ : $f_4$ is $P$ even but $C$
odd,
while $f_6$ and $f_7$ are $C$ even and $P$ odd. 
Note the slight difference in notation compared to 
\cite{Hagi};
we have chosen to write the vertex 
in a way such that the form factors $f_{i}^{V}$
express exactly the deviations of the couplings
from their SM tree level values. Indeed, by
comparing Eq.(\ref{BSMV}) with Eq.(\ref{SMV}) we see that 
all form factors $f_{i}^{V}$ are 
normalized to
be zero at tree level. 
The above form factors receive 
non-zero
contributions of order $O(\alpha)$ 
from one-loop quantum corrections within the SM 
\cite{pap,proto}; 
in fact,
$f_4^V$,$f_5^V$,$f_6^V$, and $f_7^V$ receive contributions from 
one-loop fermionic diagrams only \cite{Hagi}. 
The contributions to the $f^V_i$ obtained
from supersymmetric 
\cite{SUSY} and other
extensions of the SM \cite{extensions},
as well as composite models \cite{composite},
have been studied extensively in the 
recent literature.
In what follows we will 
treat the $f^V_i$
as if they were small with respect to unity, but not 
necessarily of order $\cal O(\alpha)$.

It is possible to impose  
constraints
on the form of the $f_i^V$ by resorting to
various physical and field-theoretical considerations.
For example,
if one requires that $\Gamma^\gamma_{\mu\alpha\beta}$ satisfies
the Ward identity of 
Eq.\ (\ref{WIOS}), namely 
$q^{\mu}\Gamma^{\gamma}_{\mu\alpha\beta}=0$, in order for
the cancellation of the gauge-dependent terms stemming
from the tree-level photon propagator 
of Eq.\ (\ref{PhotPol}) to go through as before,
the $f_4$ and $f_5$ terms in the $\gamma W^{+}W^{-}$ vertex 
$\Gamma^{\gamma}_{\mu\alpha\beta}$ need be replaced by :
\bea
-if_4^{\gamma} [q_{\beta}g_{\mu\alpha}+q_{\alpha}g_{\beta\mu}
-2 q^{\mu}q^{\alpha}q^{\beta}/q^2] 
 +if_5^{\gamma} [\epsilon^{\mu\alpha\beta\rho}(p_2-p_1)_{\rho}
 -q^{\mu}\epsilon^{\alpha\beta\rho\sigma}q^{\rho}(p_2-p_1)_{\sigma}/q^2
].
\eea
Analyticity at $q^2\!=\!0$, provides us then with the constraint 
$f_4^{\gamma}(q^2\!=\!0)\!=\!f_5^{\gamma}(q^2\!=\!0)\!=\!0$. In addition, 
fixing the electric charge of the 
$W^{\pm}$ to be $\pm 1$ imposes the additional constraint
$f_1^{\gamma}(q^2\!=\!0)\!=\!0$.

If   the form  factors  $f_i^V$   are  kept arbitrary,   the vertex of
Eq.(\ref{BSMV}) leads    to   cross-sections  which    grossly violate
unitarity, because the subtle cancellations enforced by the tree-level
couplings of the SM are now distorted \cite{CLT,CLS}.
Unitarity can only
be restored if the form factors $f_i(q^2)$ fall sufficiently fast with
increasing  $q^2$.  In fact in    analysing the effects  of  anomalous
couplings in  hadron colliders,  that probe  a wide  range of  $q^2$ a
behaviour    of  the  following    form  is  assumed   
\be f_i(q^2) = \frac{f_i^0}{(1+q^2/\Lambda^2)^2} ~.  
\ee 
The  exact form of the  form
factors depends   on the underlying  dynamics  that generate  them and
determine  the  scale $\Lambda$.  Tree   level unitarity then provides
order of magnitude estimates for the product $f_i^0 \Lambda$ \cite{Unit}
\cite{GouTsi}. 

\setcounter{equation}{0}
\section{$W$ pair production with anomalous form factors}

We now proceed to calculate the process 
\be
e^-(k_1,s_1)e^+(k_2,s_2) \rightarrow
W^-(p_1,\lambda_1)W^+(p_2,\lambda_2)
\ee
using the non-standard vertex of Eq.(\ref{BSMV}) and keeping
only terms linear  
 in the anomalous couplings 
$f_i^V$. 
For the rest of this paper we restrict ourselves only to the 
form factors that separately respect the discrete $C$,$P$ and $T$
symmetries.
The electrons are considered massless, and $s_i$, $\lambda_i$ 
label respectively 
the initial electron's and  positron's spins and the polarizations 
of the final $W$s. 
The relevant kinematical variables
in the center-of-mass frame are 
\bea
s  = & (k_1+k_2)^2  = &(p_1+p_2)^2 ~,\nonumber\\
t  = & (k_1-p_1)^2  = &(p_2-k_2)^2 = 
-\frac{s}{4}(1+\beta^2-2 \beta \cos\theta)~,
\label{defst}
\eea
where 
\be
 \beta = \sqrt{1-\frac{4M_W^2}{s}}~,
\ee
is the velocity of the $W$s, and $\theta$ 
is the
angle between the incoming electron and the outgoing $W^-$. 

\medskip
%%%%%%%%%%%%%%%%%%%%%%%
 
\begin{center}
\begin{picture}(400,130)(0,20)

\put( 34, 58){\makebox(0,0)[c]{\small $e^{+}(k_{2},s_{2})$}}
\put( 34,144){\makebox(0,0)[c]{\small $e^{-}(k_{1},s_{1})$}}
\ArrowLine( 40, 70)( 55,100)
\ArrowLine( 40,130)( 55,100)
\put( 73,115){\makebox(0,0)[c]{\small $\gamma$}}
\Photon( 55,100)( 85,100){4}{3}
\Photon( 85,100)(100, 70){4}{3}
\ArrowLine( 93, 85)( 90.5, 82)
\Photon( 85,100)(100,130){4}{3}
\ArrowLine( 91,115)( 96,115)
\put(111, 58){\makebox(0,0)[c]{\small $W^{+}(p_{2},\lambda_{2})$}}
\put(111,144){\makebox(0,0)[c]{\small $W^{-}(p_{1},\lambda_{1})$}}
\put( 70,40){\makebox(0,0)[c]{(a)}}

\ArrowLine(170, 70)(185,100)
\ArrowLine(170,130)(185,100)
\put(203,115){\makebox(0,0)[c]{\small $Z$}}
\Photon(185,100)(215,100){4}{3}
\Photon(215,100)(230, 70){4}{3}
\ArrowLine(223, 85)(220.5, 82)
\Photon(215,100)(230,130){4}{3}
\ArrowLine(221,115)(226,115)
\put(200,40){\makebox(0,0)[c]{(b)}}
 
\ArrowLine(300, 70)(330, 85)
\ArrowLine(300,130)(330,115)
\put(322,100){\makebox(0,0)[c]{\small $\nu_{e}$}}
\Line(330, 85)(330,115)
\Photon(330, 85)(360, 70){-4}{3}
\ArrowLine(345, 77.5)(347, 79)
\Photon(330,115)(360,130){4}{3}
\ArrowLine(345,122.5)(347,121)
\put(330,40){\makebox(0,0)[c]{(c)}}
 
\end{picture}
\end{center}
 
\noindent
{\small Fig.~1.
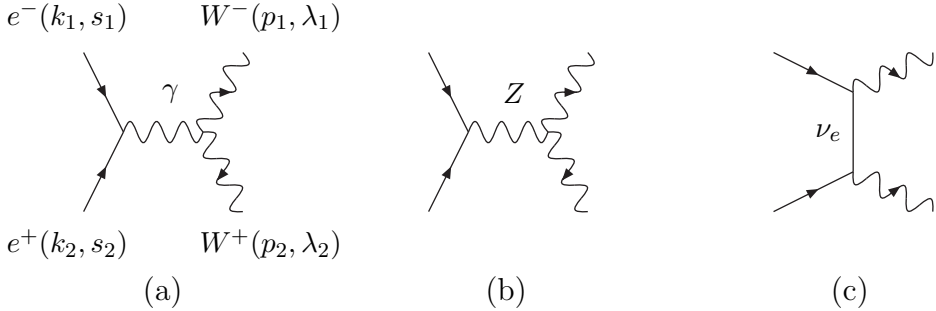
The three diagrams which contribute to the process
$e^{+}e^{-} \rightarrow W^{+}W^{-}$ at tree level in the
case of massless electrons.
\bigskip

The $S-$matrix element for this process is given by 
\be
i\langle W^{+}W^{-}|T|e^{+}e^{-}\rangle =
i\epsilon^*_{\alpha}(p_1,\lambda_1)
\epsilon^*_{\beta}(p_2,\lambda_2)\bar{v}(k_2,s_2)T^{\alpha\beta}
u(k_1,s_1)~,
\ee
where the amplitude $T^{\alpha\beta}$ is the sum 
of the graphs of Fig.1.
If we now write the vertex of Eq.(\ref{BSMV})
as a sum of the standard vertex of Eq.(\ref{SMV})
and a non-standard piece, i.e.
\be
\Gamma^V_{\mu\alpha\beta}(q,-p_1,-p_2) \equiv
\Gamma_{\mu\alpha\beta}^{V,0}(q,-p_1,-p_2) + 
\delta\Gamma^V_{\mu\alpha\beta}(q,-p_1,-p_2)~,
\ee
then
$T^{\alpha\beta}$ consists of the standard part,
$T^{0}_{\alpha\beta}$
and a non-standard part, $\delta T_{\alpha\beta}$,
which originates from the $s-$channel graphs only, i.e.
\be
T_{\alpha\beta} = T^{0}_{\alpha\beta} +\delta T_{\alpha\beta}~. 
\ee 
The two terms are given explicitly by 
\be
T^{0}_{\alpha\beta} = ig^2\left[s_w^2\gamma^{\mu} \frac{1}{s} 
+\gamma^{\mu}(v-a \gamma_5) \frac{1}{s-M_Z^2}\right] 
\Gamma_{\mu\alpha\beta}^{0}
-i\frac{g^2}{2}\gamma_{\beta}P_L\frac{1}{\ksl_1-\slp_1}
\gamma_{\alpha}P_L ~,
\ee 
and 
\be
\delta T_{\alpha\beta} = ig^2s_w^2\gamma^{\mu} \frac{1}{s}
\delta\Gamma^{\gamma}_{\mu\alpha\beta}
+\gamma^{\mu}(v-a \gamma_5) \frac{1}{s-M_Z^2} 
\delta\Gamma^Z_{\mu\alpha\beta}~,
\ee
In the above equations 
 $P_L=(1-\gamma_5)/2$ is the left chirality projector 
and $v=1/4-s^2_w$,
$a=1/4$ 
are respectively 
the vector and axial couplings of the electron with the $Z$.

Defining the $W$ polarization tensor 
as
$Q_{\mu\nu}(k) \equiv \sum \epsilon_{\mu}(k)\epsilon_{\nu}^{*}(k)
= -g_{\mu\nu}+ k_{\mu}k_{\nu}/M^2_W $~,
the modulus squared of the 
matrix element averaged over initial state spins, summed   
over the final polarizations, and to first order in the deviations is 
given by
\bea\label{TT}
\sum_{s_{1},s_{2}}\sum_{\lambda_{1},\lambda_{2}}
|\langle e^{+}e^{-}|T|W^{+}W^{-}\rangle|^{2}
&=& \sum_{s_{1},s_{2}}
\biggl(\overline{v}\,T_{\mu'\nu'}u\biggr)^{*}
Q^{\mu'\mu}(p_1)Q^{\nu'\nu}(p_2)
\biggl(\overline{v}\,T_{\mu\nu}u\biggr)\nonumber\\
&=&\sum_{s_{1},s_{2}}
\biggl(\overline{v}\,T_{\mu'\nu'}^{0}u\biggr)^{*}
Q^{\mu'\mu}(p_1)Q^{\nu'\nu}(p_2)
\biggl(\overline{v}\,T_{\mu\nu}^{0}u\biggr)\nonumber\\
&+& 2\sum_{s_{1},s_{2}}
\Re e \Big[\biggl(\overline{v}\,T_{\mu'\nu'}^{0}u\biggr)^{*}
Q^{\mu'\mu}(p_1)Q^{\nu'\nu}(p_2)
\biggl(\overline{v}\,\delta T_{\mu\nu}u\biggr)\Big]~.
\eea

The unpolarized differential cross-section in the center-of-mass
frame is given by
\be 
\frac{d\sigma}{d\Omega}= \frac{1}{256\pi^2}\frac{\beta}{s}
\sum_{s_{1},s_{2}}\sum_{\lambda_{1},\lambda_{2}}
|\langle e^{+}e^{-}|T|W^{+}W^{-}\rangle|^{2}~,
\ee
and consists of a standard and an anomalous contribution
\be
\frac{d\sigma}{d\Omega} = 
   \frac{d\sigma^{0}}{d\Omega}+\frac{d\sigma^{an}}{d\Omega}~,
\label{CS}   
\ee
given by
\be
\frac{d\sigma^{0}}{d\Omega}
= \frac{1}{256\pi^2} \frac{\beta}{s}
\sum_{s_{1},s_{2}}
\biggl(\overline{v}\,T_{\mu'\nu'}^{0}u\biggr)^{*}
Q^{\mu'\mu}(p_1)Q^{\nu'\nu}(p_2)
\biggl(\overline{v}\,T_{\mu\nu}^{0}u\biggr)~,
\label{SCS}
\ee
and
\be
\frac{d\sigma^{an}}{d\Omega} = \frac{1}{128\pi^2}\frac{\beta}{s}
 \sum_{s_{1},s_{2}}
\Re e\Big[\biggl(\overline{v}\,T_{\mu'\nu'}^{0}u\biggr)^{*}
Q^{\mu'\mu}(p_1)Q^{\nu'\nu}(p_2)
\biggl(\overline{v}\,\delta T_{\mu\nu}u\biggr)\Big]~, 
\label{ANCS}
\ee
respectively.
The SM cross-section $d\sigma^{0}/d\Omega$
in Eq.(\ref{SCS})
has been calculated long 
time ago in \cite{Buras}; we do not report it here. 
We next turn to the computation
of the non-standard part of the cross-section 
$d\sigma^{an}/d\Omega $.
We  define the variables  
\be 
x=\cos\theta, ~~~~~ z=\frac{1+\beta^2}{2\beta}~,
\ee
in terms of which the Mandelstam variable $t$, 
defined
in Eq.(\ref{defst}),
becomes 
\be
t=-\frac{s}{4}(1+\beta^2-2\beta x ) = -\frac{s\beta}{2}(z-x)~.
\ee
Then, a straightforward calculation yields:
\be
(z-x)\frac{d\sigma^{an}}{d\Omega} = \frac{1}{128\pi^2}
g^4\frac{\beta}{s}\sum_{i=1}^{4}\sigma_i(s)P_i(x,s)~.
\label{PolExp}
\ee
The functions $P_i(x,s)$ are polynomials in $x$ of maximum degree 3, 
given by 
\bea
P_1(x,s)& =& z-x ~,\nonumber\\
P_2(x,s)& =& (z-x)(1- x^2) ~,\nonumber\\
P_3(x,s)& =&  1- x^2 ~,\nonumber\\
P_4(x,s)& =&  1- \beta  x ~.
\eea

In  Eq.(\ref{PolExp}) the terms proportional to
$P_1$ and $P_2$ 
originate from 
the square of the $s-$channel graphs,
whereas the terms proportional to     
 $P_3$ and $P_4$ from the interference between the $s-$channel graphs
 and the $t$- channel graph.
Notice that there is no contribution to 
 $d\sigma^{an}/d\Omega$ originating from
the square of the $t$- channel graph; this is so
because all such contributions are absorbed into the
standard part, since the couplings of the
neutral gauge bosons to fermions  
are assumed to be exactly those of the SM.
 
The functions $\sigma_i(s)$ are linear combinations of the 
various deviation form factors. Specifically:
\bea
\sigma_1(s) &=& A_1 f_3^{\gamma} + A_2 f_3^Z ~,\nonumber\\
\sigma_2(s) &=& A_3 f_1^{\gamma} + A_4 f_1^Z +A_5 f_3^{\gamma} 
+ A_6 f_3^Z + A_7 f_2^{\gamma} + A_8 f_2^Z ~,\nonumber\\
\sigma_3(s) &=& A_9 f_1^{\gamma} + A_{10} f_1^Z -\eta A_9 f_2^{\gamma}  
-\eta A_{10} f_2^Z ~,\nonumber\\
\sigma_4(s) &=& A_{11} f_3^{\gamma} + A_{12} f_3^Z~, 
\label{sigma}
\eea
where $\eta = s/4M_W^2$.
The explicit closed 
expressions for the coefficients $A_i$ are given below.
Setting
\bea
y &=& \frac{s}{s-M_Z^2}\frac{1}{c_w^{2}} ~, \nonumber\\
r &=&  a^2+v^2~,
\eea
we have
\bea
A_1 &=&  -8 s_w^2 \left[y v(4c_w^2-1) + 4 s_w^2 \right] ~,
\nonumber  \\
A_2 & =&-8yc_w^2\left[yr(4c_w^2-1)+4s_w^2 v \right]~, 
\nonumber\\
A_3 &=&  \beta^2 s_w^2  
[2(3+2\eta)(y v c_w^2+s_w^2) -(1+2\eta)y v]~, \nonumber \\
A_4 & = & \beta^2yc_w^2
\left[2(3+2\eta)( yrc_w^2+vs_w^2 )  
-(1+2\eta)y r\right]~,
\nonumber\\
A_5 & = & -4\beta^2\eta s_w^2 [yv(2c_w^2 -1)+2s_w^2]~,     
\nonumber\\
A_6 & = & - 4\beta^2\eta y c_w^2
[yr(2c_w^2 -1)+2v s_w^2]~,   
\nonumber\\
A_7 & = &  -2\beta^2\eta s_w^2
[2(1+\eta)(y vc_w^2+ s_w^2) -\eta y v]~, \nonumber\\
A_8 &= & -2\beta^2\eta yc_w^2 
[2(1+\eta)(y r c_w^2+  v s_w^2)
-\eta y r]~,    
\nonumber\\
A_9  &=& -\beta s_w^2 ~,\nonumber \\
A_{10} &=& -(a+v)\beta y c_w^2 ~,\nonumber \\
%% A_{11}  &=& s_w^2\eta  ~, \nonumber \\
%% A_{12}  &=& (a+v)y\eta  ~, \nonumber \\
 A_{11} &=& 4s^2_w\beta^{-1} ~,\nonumber\\
 A_{12} &=&  4 y(a+v)c_w^2\beta^{-1}~.  
\eea

The values of the coefficients $A_i(s)$ for some typical 
LEP2 energies are shown in Table 1.
Notice that the 
coefficients $A_i$ 
appearing within each of the 
four equations in (\ref{sigma})
are  
of the same order of magnitude, and therefore none of them can be neglected.

\medskip

\begin{center}
\begin{tabular}{|c|l|l|l|l|l|} \hline
$\sqrt{s}$ (GeV)& ~~~~161 & ~~~~172  & ~~180  & ~~192   &~~200 \\ \hline\hline
$A_1$    &-1.79    &-1.78   &-1.77  &-1.77  &-1.76  \\ \hline
$A_2$    &-3.25    &-2.91   &-2.73  &-2.53  &-2.42  \\  \hline\hline
$A_3$    &~2.20$\times10^{-3}$ &~7.40$\times10^{-2}$ &~0.123 &~0.191 &~0.235 \\  \hline
$A_4$    &~3.68$\times10^{-3}$ &11.0~$\times10^{-2}$  &~0.169 &~0.239 & 0.279 \\ \hline
$A_5$    &-1.70$\times10^{-3}$ &-6.18$\times10^{-2}$ &-0.108 &-0.181 &-0.232 \\\hline 
$A_6$    &-1.83$\times10^{-3}$ &-6.03$\times10^{-2}$ &-0.0991&-0.154 &-0.190 \\ \hline
$A_7$    &-1.79$\times10^{-3}$ &-6.94$\times10^{-2}$ &-0.127 &-0.228 &-0.306 \\ \hline
$A_8$    &-3.24$\times10^{-3}$ &-11.1$\times10^{-2}$  &-0.187 &-0.303 &-0.385 \\ \hline\hline
$A_9$    &-1.41$\times10^{-2}$ &-7.98$\times10^{-2}$ &-0.101 &-0.122 &-0.133 \\ \hline
$A_{10}$ &-2.56$\times10^{-2}$ &-13.7$\times10^{-2}$ &-0.167 &-0.195 &-0.208 \\ \hline\hline
$A_{11}$ &14.2     &~2.51   &~1.99  &~1.64  &~1.50  \\ \hline
$A_{12}$ &25.8     &~4.31   &~3.30  &~2.61  &~2.34  \\  \hline
\end{tabular}
\end{center}

\noindent {\bf Table 1} : The coefficients $A_i$ as a function of $s$. 
\vspace*{0.5cm}
%%%%%%%%%%%%%%%%%%%%%%%%%%%%%%%%%%%%%%%%%%%%%%%%%%%%%%%%%%%%%%%%%%%%%%%%%%%

The polynomials $P_i(x)$ are
linearly independent; indeed, their Wronskian
is given by 
\be
W(P_i) = \frac{24M^2_W}{s}~,
\ee 
which can be  zero only if $s\to \infty$.

It is important to emphasize at this point
that in our case the quantity 
$(z-x)({d\sigma^{an}}/{d\Omega})$ appearing on the
right hand side of Eq.(\ref{PolExp}) is more suitable
as an experimental observable than
$({d\sigma^{an}}/{d\Omega})$ itself \cite{PRW}.
The reason is that this quantity is the sum 
of a finite number of linearly independent  polynomials.
Instead,
an expansion of $({d\sigma^{an}}/{d\Omega})$
in terms of polynomials in $x$ would necessitate
an infinite number of them, because of the
presence of the term $(z-x)^{-1}$. 
This fact would in turn complicate
the inversion of such a relation, 
i.e. the determination of the quantities 
$\sigma_i$, which contain the dependence on
the $f_{i}^{V}$.
We next 
proceed to carry this inversion for 
the quantity $(z-x)({d\sigma^{an}}/{d\Omega})$.

To accomplish this
one must 
construct
a set of four other polynomials, $\widetilde{P}_i(x)$, which are
orthonormal 
to the $P_i(x)$, i.e. they satisfy 
\be
\int_{-1}^{1}\widetilde{P}_i(x,s)P_j(x,s) dx = \delta_{ij} ~.
\ee
These polynomials are: 
\bea
\widetilde{P}_1(x,s)& =&%\frac{1}{4(1-\beta z)}
\frac{1}{8\eta}
(3 \beta +15x - 15 \beta x^2
 - 35 x^3) ~, \nonumber\\
\widetilde{P}_2(x,s)& =& \frac{35}{8} (-3x+5x^3) ~,\nonumber\\
\widetilde{P}_3(x,s)& =& \frac{5}{8}(3 +21x z -9x^2-35x^3 z) ~, \nonumber\\
\widetilde{P}_4(x,s)& =&%\frac{1}{4(1-\beta z)}
\frac{1}{8\eta}(-3  -15 x z + 15 x^2 + 35 x^3
z )~.
\eea
Notice that $\widetilde{P}_2(x)$ is proportional to the 
third Legendre polynomial.

Thus, the $\sigma_i$ are given by
\be
\sigma_i = \left[\frac{64\pi s}{g^4\beta}\right] 
 \int_{-1}^{1} dx (z-x)  
(\frac{d\sigma^{an}}{dx})\widetilde{P}_i(x,s) ~,
\label{SIG}
\ee
where the trivial $d\phi$ integration has been carried out.

If we assume that 
the experimental value $d\sigma^{exp}/d\Omega$
for the differential cross-section of the above process
has been measured, and that
all new physics is parametrized by non-standard 
trilinear vector couplings, then we have 
that 
\be
\frac{d\sigma^{an}_{exp}}{d\Omega} = 
\frac{d\sigma_{exp}}{d\Omega}-   
\frac{d\sigma^{0}}{d\Omega} ~.
\label{crosassum}
\ee
Therefore, the experimental values for 
$\sigma_i$ are given by 
\be
\sigma_i^{exp} = \left[\frac{64\pi s}{g^4\beta}\right] 
 \int_{-1}^{1} dx (z-x)  
\left(\frac{d\sigma^{exp}}{dx}-\frac{d\sigma^{0}}{dx}\right)\widetilde{P}_i(x,s)
~.
\label{SIGEXP}
\ee

The fact that one can extract, at least in principle,
experimental information for the quantities $\sigma_i$
motivates the study of the system of equations given in 
Eq.(\ref{sigma}). 
Of course,
since Eq.(\ref{sigma})
constitutes a system
of four equations for six unknown quantities,
we do not expect to determine all
deviation form factors $f_i^V$ individually.
We can easily do so however for two of them;
indeed 
the first and fourth equations 
in (\ref{sigma})
constitute a separate
system of two equations with two unknowns, $f^{\gamma}_{3}$ 
and $f^Z_{3}$, 
which can be solved exactly:
\bea 
f_3^{\gamma} & = &  \gamma_1\sigma_1+\gamma_4\sigma_4 ~,
\nonumber\\
f_3^Z  & = & z_1\sigma_1+z_4\sigma_4 ~,
\label{anmagmom}
\eea 
where  
\bea
\gamma_1 & = & 
-\frac{a+v}{8as^4_w}\cdot\frac{1}{[y(1-4 c_w^2)+4]} ~,\nonumber  \\
\gamma_4 & = & 
-\frac{1}{4as^4_w}\cdot\frac{\beta[y r (4 c_w^2-1)+4 s_w^2 v]}
{[y(1-4 c_w^2)+4]}~,\nonumber\\
z_1 & = & \frac{1}{8 s_w^2 c_w^2 a}\cdot\frac{1}
{ y [y(1-4 c_w^2)+4]}~,\nonumber\\
z_4 & = & \frac{1}{4 s_w^2 c_w^2 a}\cdot
\frac{\beta[y v (4 c_w^2-1)+4 s_w^2 ]}{y[y(1-4 c_w^2)+4]}~.
\eea
Thus, the measurement of the two observables 
$\sigma_1$ and $\sigma_4$ {\sl directly} 
determines the deviations %of $f_3^{\gamma}$ and $f_3^Z$ 
from their SM values, of the magnetic dipole form factors 
$G_M^{\gamma}$ and $G_M^{Z}$
of the $W$ due to the photon and the $Z$ respectively :  
\bea
G_M^{\gamma}(s) &=& \frac{e}{2M_W}\left[2+f_3^{\gamma}\right] =
\frac{e}{2M_W}\left[2+\gamma_1\sigma_1+\gamma_4\sigma_4\right]~,
\nonumber\\
G_M^{Z}(s) &=& \frac{e}{2M_W}\frac{c_w}{s_w}\left[2+f_3^{Z}\right] =
\frac{e}{2M_W}\frac{c_w}{s_w}\left[2+z_1\sigma_1+z_4\sigma_4\right]~.
\eea

We are not
aware of the existence in the literature of similar simple expressions
relating directly the anomalous magnetic moments to 
the {\it unpolarized} differential cross-section with two on-shell
$W$ bosons.

The remaining two equations provide
two constraints between four of the 
six non standard form factors. In particular
\bea  
a_1 \sigma_1+ \sigma_2 + a_4 \sigma_4&=&
A_3 f_1^{\gamma} + A_4 f_1^Z+ A_7 f_2^{\gamma} + A_8 f_2^Z 
\ , \nonumber\\
\sigma_3 &=& A_9 f_1^{\gamma} + A_{10} f_1^Z -\eta A_9 f_2^{\gamma}  
-\eta A_{10} f_2^Z \ ,
\label{sumrule}
\eea
where
\bea
a_1 &=&  -\frac{\beta^2 \eta}{2}\frac{y(1-2 c^2_w)+2}{y(1-4 c^2_w)+4}
~,\nonumber\\
a_4 &=& -\frac{2 a\beta^3\eta y }{y(1-4 c^2_w)+4}~,
\eea
and we have used that $a-v=s_w^2$. The above equations can be cast in 
the equivalent form of two sum rules \cite{SumRules}
\bea
 \int_{-1}^{1} dx (z-x)  
\left(\frac{d\sigma^{exp}}{dx}-\frac{d\sigma^{0}}{dx}\right)
H(x,s)&=& 
A_3 f_1^{\gamma} + A_4 f_1^Z+ A_7 f_2^{\gamma} + A_8 f_2^Z 
\ , \nonumber\\
\int_{-1}^{1} dx (z-x)  
\left(\frac{d\sigma^{exp}}{dx}-\frac{d\sigma^{0}}{dx}\right)
\widetilde{P}_3(x,s) &=& 
A_9 f_1^{\gamma} + A_{10} f_1^Z -\eta A_9 f_2^{\gamma}  
-\eta A_{10} f_2^Z \ ,
\label{SumRule}
\eea
where 
\be
H(x,s) = a_1\widetilde{P}_1(x,s)+\widetilde{P}_2(x,s)
+a_4\widetilde{P}_4(x,s) \ .
\ee

The analysis and the results presented 
thus far are model-independent
since no assumptions have been made about the 
dynamical mechanism which gives rise to the anomalous
couplings. It would be interesting to examine how
the above results could be used for testing the
validity of specific models which predict the
 generation of such couplings.
We will discuss some of these issues in the next section,
in the context of a composite model for the $W$.

\setcounter{equation}{0}
\section{Predictions of a composite model} 

In this section we examine the predictions at the level of
experimental cross-sections
of a model of compositeness 
proposed by Brodsky and Hiller \cite{BH}. 
In this model the $W$ is considered a bound state of two,
in general different, fermions. The 
two fermions are held together by the exchange of a
 gauge boson of mass $\lambda$. The results presented
have been 
calculated in the one-boson exchange approximation
(Fig.2); as explained in \cite{BH} the model becomes
gauge-invariant in the collinear approximation.

The matrix element 
$(G_{h,h'}^{V})_{\mu}= <p_2,h'|J^{\mu}_V |p_1,h>$
of the current $J^{\mu}_V$
between
the momentum and helicity eigenstates $|p_1,h>$ and $|p_2,h'>$
is written in terms of three-form factors
\footnote{
We use the same notation as in \cite{BH}, except for the 
labelling of the four-momenta, where we have set 
$p'\to p_2$ and $p\to -p_1$}
as follows:
\bea
(G_{h,h'}^{V})_{\mu}&=& -G_1^V(q^2) ({\epsilon'}^{*}\cdot\epsilon)
( p_2-p_1)_{\mu} \nonumber\\
&& - G_2^V(q^2)
[({\epsilon'}^{*}\cdot q)\epsilon_{\mu}
- (\epsilon\cdot q) {\epsilon'}^{*}_{\mu}]\nonumber\\
&& + G_3^V(q^2)/2M_W^2
({\epsilon'}^{*}\cdot q) (\epsilon \cdot q) (p_2-p_1)_{\mu}
\eea
where $\epsilon \equiv \epsilon_{h}$ and $\epsilon'\equiv \epsilon_{h'}$
are the initial and final polarisation vectors.

The kinematical form factors $ G^V_1$, $G^V_2$, and $G^V_3$ are related to the 
photonic ($V=\gamma$), or weak ($V=Z$)  
charge $G^V_C$, magnetic  dipole $G^V_M$,
 and electric  quadruple $G^V_Q$ 
form factors of the $W$ 
through the relations:
\bea 
G^V_C& =& G_1^V +\frac{2\eta}{3}G^V_Q ~,\nonumber\\ 
G^V_M& =& G_2^V~\nonumber \\
G^V_Q& =& G_1^V -G_2^V +(1+\eta)G_3^V
\eea

As was shown in \cite{BH},
this model  
predicts the following ratios for photon form factors
\be
G_C^{\gamma}:G_M^{\gamma}:G_Q^{\gamma}=(1-\frac{2}{3}\eta):2:-1
\ee
which, in the notation of Eq.(1), translates into 
\be
f_2^{\gamma}=0,~~~~ f_1^{\gamma} =f_3^{\gamma}~.
\label{compcon}
\ee
Notice that the above ratios, which have been 
derived using the non-trivial dynamics
of this composite model, coincide with  
the corresponding ratios satisfied by the 
tree-level SM values of $G_C$, $G_M$, and $G_Q$.

However, no analogous 
ratios for the 
form factors  
$G_C^{Z}$, $G_M^{Z}$, and $G_Q^{Z}$ 
were derived in \cite{BH}, because
their analysis had focused on the photon form-factors only. 
In what follows we will address this issue in some detail. 
In particular we will study 
how the anomalous gauge boson
couplings
predicted by this model 
are affected by the coupling of the 
neutral gauge bosons $V$, ($V=\gamma, Z$) to the fermions,
which make up the composite $W$.

\pic

\vspace*{3.5cm}

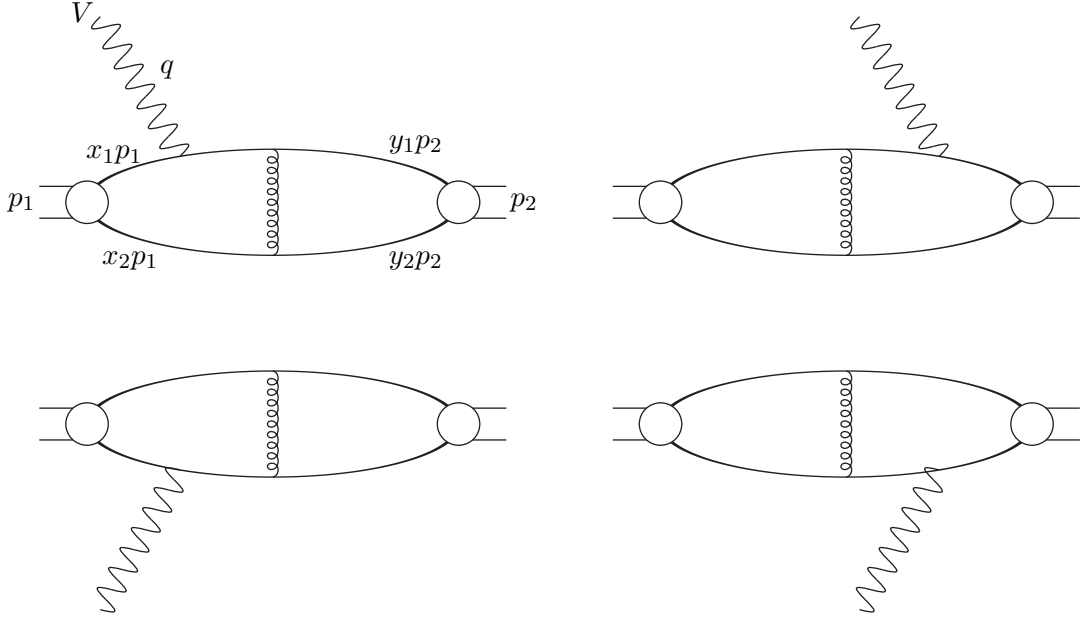
\noindent {\small Fig.~2 The composite model in the one-boson exchange
approximation}

\vspace*{1cm}
 
The most general coupling of the $i$-th fermion ($i=1,2$) 
allowed by Lorentz invariance 
is given by
\be
\Gamma_{\mu}^{V,i}=\gamma_{\mu}F_{1}^{V,i}
+\sigma_{\mu\nu}q^{\nu}F_{2}^{V,i}
+\gamma_{\mu}\gamma_{5}F_{3}^{V,i}+\sigma_{\mu\nu}q^{\nu}\gamma_{5}
F_{4}^{V,i}~,
\label{GFV}
\ee
where $\sigma_{\mu\nu}=\frac{i}{2}[\gamma_{\mu},\gamma_{\nu}]$.
Note that $F_{1}^{V,i}$ and $F_{3}^{V,i}$ are dimensionless
quantities, whereas $F_{2}^{V,i}$ and $F_{4}^{V,i}$ have dimensions
of inverse mass.
To determine the effect 
of this general fermion-boson 
coupling 
on the anomalous trilinear gauge boson couplings
we must repeat the calculation presented in \cite{BH},
keeping the general form for $\Gamma_{\mu}^{V,i}$
given on the left hand side of Eq.({\ref{GFV}), instead of only the term
proportional to $\gamma_{\mu}$.

From \cite{BH} we know that
\be
(G_{h,h'}^{V})_{\mu} \sim \frac{1}{s}\frac{1}{(s-4\lambda^2)}
\sum_{i=1,2}[(A_{hh'}^{V,i})_{\mu}+(B_{hh'}^{V,i})_{\mu}]~,
\ee 
with
\bea
(A^{V,i}_{hh'})_{\mu} &=&
{\mbox Tr}\{\gamma_{\nu}\bar{\chi}_{Jh'}\gamma^{\nu}
[\not\! p_2-\frac{1}{2}(\not\! p_1-M)]\Gamma_{\mu}^{V,i}{\chi}_{Jh}\}~,
\nonumber\\
(B^{V,i}_{hh'})_{\mu} &=&
-{\mbox Tr}
\{\bar{\chi}_{Jh'}\Gamma_{\mu}^{V,i}[\not\! p_1 +\frac{1}{2}(\not\! p_2-M)]
\gamma^{\nu}{\chi}_{Jh}\gamma_{\nu}\}~,
\eea
where $M$ is the mass of the composite $W$, and
the spin wave functions are given by
\be
{\chi}_{1h}=\frac{-1}{\sqrt{2}}\not\! \epsilon_{h}(\not\! p -M)~~,
~~~~~~~
{\chi}_{00}=\frac{1}{\sqrt{2}}\gamma_5 (\not\! p -M)~.
\ee

If we define the quantity
\be
{\tilde F}_j^V= \sum_{i} F_{j}^{V,i} ~~,  ~~j=1,2,3,4
\ee
we find 
in the limit where the mass $M$ of the composite $W$
and the masses $m_1$ and $m_2$ of the constituent fermions
satisfy $M=m_1+m_2$
($x_1=x_2=y_1=y_2=1/2$ in the notation of \cite{BH}) :
\bea
f_1^V &=& {\tilde F}_1^V-\frac{2s}{M}{\tilde F}_2^V-1~,  \nonumber\\ 
f_2^V &=& -4M{\tilde F}_2^V ~,\nonumber\\
f_3^V &=& {\tilde F}_1^V ~
-\left(\frac{s+4M^2}{2M}\right){\tilde F}_2^V-1 ~, \nonumber\\
f_4^V &=&0 ~, \nonumber\\
f_5^V &=& -i{\tilde F}_3^V +i\left(\frac{s}{2M}\right) {\tilde F}_4^V ~, 
\nonumber\\
f_6^V &=& 4M{\tilde F}_4^V ~,\nonumber\\
f_7^V &=& 2M{\tilde F}_4^V ~.
\eea
In deriving these results we have made use of the identities
listed in Eq.(A3) and Eq.(A4) in the Appendix of \cite{Hagi}.
We see that the presence of ${\tilde F}_2^V$ distorts the 
compositeness condition, whereas ${\tilde F}_3^V$ and ${\tilde F}_4^V$ 
do not enter in the definition of $f_1^V$, $f_2^V$ and $f_3^V$.

Having established the above results we will now pursue two different
possibilities: (i) we will assume that the compositeness
condition holds for the photonic form factors only, i.e.   
${\tilde F}_2^{\gamma}=0$ but ${\tilde F}_2^Z \neq 0$
(ii) we will assume that the compositeness condition holds for both the
photonic and $Z$ form factors, i.e. ${\tilde F}_2^{\gamma}={\tilde F}_2^Z =0$.

In the first case, using 
Eq {\ref{compcon}, which we assume to be true, we
obtain from Eq.\ (\ref{anmagmom}) and 
Eq.\ (\ref{sumrule}) the following system of two equations for
the two remaining  unknown form factors $f_1^Z$, $f_2^Z$:
\bea
(a_1-A_3\gamma_1)\sigma_1+\sigma_2+(a_4-A_3\gamma_4)\sigma_4&=&
A_4 f_1^Z+ A_8 f_2^Z  
\ , \nonumber\\
-A_9\gamma_1\sigma_1+\sigma_2-A_9\gamma_4\sigma_4&=&
A_{10} f_1^Z -\eta A_{10} f_2^Z \ ,
\label{sumrulec1}
\eea
This yields the solutions 
\bea
f_1^Z&=&
\sigma_1(\gamma_1 b_1-\eta a_1 A_{10})/D_1
-\sigma_2\eta  A_{10}/D_1
-\sigma_3A_{8}/D_1
+\sigma_4(\gamma_4 b_1-\eta a_4 A_{10})/D_1~,
\nonumber\\
f_2^Z&=&
\sigma_1(\gamma_1 b_2- a_1 A_{10})/D_1
-  \sigma_2A_{10}/D_1
+\sigma_3A_{4}/D_1
+\sigma_4(\gamma_4 b_2-a_4 A_{10})/D_1~,
\eea
with
\bea 
b_1&=& \eta  A_{10} A_{3}+A_8 A_9~,\nonumber\\
b_2&=&   A_{10} A_{3}-A_4 A_9~,\nonumber\\
D_1&=& (a+v)\beta^3\eta y^2 c_w^4 [yr(2c_w^2-1)+2 v s^2_w]~.
\eea

Thus, assuming that the couplings of $W$ to the photon obey their
tree-level SM relations one can extract 
all the remaining of the $C$ and $P$ 
conserving anomalous form-factors directly
from the differential cross section. 

In the second possibility of interest,
using the fact that,  because of the compositeness conditions
$f_2^V=0$, Eq.(\ref{sumrule}) reduces to
\bea  
a_1\sigma_1 +\sigma_2 +a_4\sigma_4 &=&
A_3 f_1^{\gamma} + A_4 f_1^Z \ ,\nonumber\\
\sigma_3 &=& A_9 f_1^{\gamma} + A_{10} f_1^Z \ .
\label{sumrule2}
\eea
Then, since $f_1^{V}=f_3^{V}$, we
 can substitute into Eq.(\ref{sumrule2})
the solutions for $f_3^{V}$ from Eq.(\ref{anmagmom})
to arrive at
\bea
\sigma_2 + c_1\sigma_1+c_2\sigma_4 & = & 0\ ,\nonumber\\
\sigma_3 + c_3\sigma_4 & = & 0\ ,
\label{predict}
\eea
where
\bea
c_1 & = &\frac{1}{8}\beta^2 \frac{2(3-2\eta)(1-yc_w^2)+(1-2\eta)y}
{4(1-yc_w^2)+y} \ ,\nonumber\\
c_2 & = &-\frac{a}{2}\beta^3 y \frac{(1+6\eta)}{4(1-yc_w^2)+y} 
\ ,\nonumber\\
c_3 & = & \frac{\beta^2}{4}
\ .
\label{sigma2}
\eea

Using   
Eq.(\ref{sigma2}) and assuming Eq.(\ref{crosassum})
we obtain the following two {\sl predictions}:
\be
 \int_{-1}^{1} dx (z-x)  
\left(\frac{d\sigma^{exp}}{dx}-\frac{d\sigma^{0}}{dx}\right)
H_i(x,s) = 0~, ~~~~~~i=1,2 
\label{Pred}
\ee
where $H_1$ and $H_2$ are polynomials in $x$ of maximum
degree three,
given by
\bea
H_1 &=& \widetilde{P}_2 + c_1 \widetilde{P}_1 + c_2 \widetilde{P}_4 ~,
\nonumber\\
H_2 &=& \widetilde{P}_3 + c_3 \widetilde{P}_4~.
\eea 

\setcounter{equation}{0}
\section{Predictions of an effective Lagrangian approach}

 We now turn our attention to an effective Lagrangian 
 approach to anomalous gauge couplings based on 
 $SU(2)_L\times U(1)_Y$ gauge invariance. In such an approach  the
deviations
 of the gauge couplings arise from gauge invariant but 
 non-standard interaction terms of dimension $d>4$,
 between
 gauge bosons and the Higgs field
  \cite{Schil,Ruj,HISZ,effthe,Goun}. 
  Such terms are assumed to originate
  from an as yet unknown underlying theory 
  at a new-physics mass scale $\Lambda$. Thus, the corresponding 
  strengths  of these interactions will be 
  suppressed by factors $(\Lambda)^{4-d}$.
  
  In order for all form factors of the vertex of Eq.(\ref{BSMV}) to be
  generated one must consider a host of operators of dimension $d$~ up
  to twelve;  in such a scenario  no constraints can be obtained among
  the various form factors. If on  the other hand the new-physics mass
  scale $\Lambda$ is  assumed to be large, e.g.  $\Lambda\ge 1$ TeV, a
  low energy  approximation where only  operators of dimension six are
  retained  will   lead  to  relations  among   the  various otherwise
  unrelated   form factors of  the vertex.   Such relations were first
  derived by imposing global $SU(2)$  symmetry on the phenomenological
  Lagrangian   that   generates     the  vertex   of   Eq.(\ref{BSMV})
  \cite{Schil,Schil2}.  In fact, the precise nature of these relations
  depends on  whether   a linear  or   non-linear  realization of  the
  symmetry is adopted. For a relatively light Higgs the new physics is
  described by   a linear realization  of  the symmetry, while   for a
  sufficiently heavy one a non-linear realization is required.
 
  We consider   first a linear realization  of  the Higgs sector.  The
  basic  ingredients of   an  $SU(2)_L\times U(1)_Y$ gauge   invariant
  Lagrangian%inducing self interactions among the gauge bosons, 
  are the Higgs field $\Phi$,  its covariant derivative $D_{\mu}\Phi$,
  and   the  non-Abelian  field   strengths   $B_{\mu\nu}$   and ${\bf
    W}_{\mu\nu}$  of   the    $U(1)_Y$ and  $SU(2)_L$   gauge  fields,
  respectively.  If one imposes the additional requirement of separate
  $C$ and $P$  invariance, there are  three operators of dimension six
  that can induce trilinear gauge couplings. They are described by the
  effective   Lagrangian  
\bea     
{\cal   L}_{eff}^{d=6}  &=&     ig'
  \frac{\alpha_{B\phi}}{M_w^2}(D_{\mu})^{\dagger} B^{\mu\nu} (D_{\nu})
  +ig \frac{\alpha_{W\phi}}{M_w^2} (D_{\mu})^{\dagger} {\bf \tau \cdot
    W}^{\mu\nu} (D_{\nu})
  \nonumber \\
  &+   &  g   \frac{\alpha_{W}}{6   M_w^2}  {\bf  W^{\mu}_{\nu}  \cdot
    (W^{\nu}_{\rho}\times W^{\rho}_{\mu})}~,
\label{d6}
\eea
where $g$ and $g'$ are the $SU(2)_L$ and $U(1)_Y$ gauge couplings, 
respectively.

The part of the above Lagrangian describing the self interactions of the 
gauge bosons is obtained by replacing 
the Higgs field with its vacuum expectation value, 
$\Phi^T\rightarrow (0,u/\sqrt{2})$. Explicitly  
\bea
{\cal L}^{WWV}_{eff} &=& ig_V\left\{ g_1^V(W^+_{\mu\nu}W^{- \mu}-
W^{+ \mu}W^-_{\mu\nu})V^{\nu} + \kappa_V W^+_{\mu}W^-_{\nu}V^{\mu\nu} 
+\frac{\lambda_V}{M_w^2} W^{+ \nu}_{\mu}W^{- \rho}_{\nu}V^{\mu}_{\rho}
\right\}~.
\label{CPeff}
\eea

In the SM the couplings $g_1^V,~\kappa_V$ and $\lambda_V$ 
have the values $g_1^V=\kappa_V=1$, $\lambda_V=0$, 
and are directly related to the 
charge, magnetic dipole, and electric quadruple moments 
of the $W$. 
$g_1^{\gamma}$ is fixed to 1 by electromagnetic gauge invariance, while 
the rest 
of the couplings  are parametrized by the available free parameters
 of the effective Lagrangian $\alpha_{W\phi}$, ~$\alpha_{B\phi}$, 
 and $\alpha_{W}$ according to :
\bea
\Delta g_1^{Z}   &=& \alpha_{W\phi}/c_w^2 ~,\nonumber\\
\Delta \kappa_{\gamma} &=& \alpha_{W\phi} +\alpha_{B\phi}~,\nonumber\\
\Delta \kappa_{Z} &=&  \alpha_{W\phi} 
-\frac{s_w^2}{c_w^2}\alpha_{B\phi}~, \nonumber\\
\lambda_{\gamma} &=& \alpha_{W}~,\nonumber\\
\lambda_{Z}  &=& \alpha_{W}~.
\label{alphas}
\eea
Since no operators containing derivatives of the form
$\Box^n V^{\mu}$ ( or $ \Box^n W^{\alpha}$  for off-shell $W$s) 
are present in Eq.(\ref{CPeff}) the resulting form factors will be 
strictly constants, independent of $s$.
Nevertheless they must scale as $M_W^2/\Lambda^2$.
The additional requirement of tree level 
unitarity imposes bounds on the products of 
the $\alpha_i$ with the  scale $\Lambda$ 
\cite{Unit,GouTsi}.

The relations of Eq.(\ref{alphas})
constitute the constraints  predicted by this model 
for the anomalous gauge couplings. They may also be found in the 
literature  in the following form :
\bea
\Delta \kappa_{Z} &=&\Delta g_1^{Z}-\frac{s_w^2}{c_w^2}\Delta \kappa_{\gamma}
~,\nonumber\\
\lambda_{\gamma} &=&\lambda_{Z}~.
\eea

In order to translate them into relations 
 among the $f_i$ form-factors, we must use that \cite{Hagi}
\bea
f_1^V &=&\Delta  g_1^V + \frac{s}{2 m_w^2}\lambda_V ~,
\nonumber\\
f_2^V &=& 2\lambda_V ~,
\nonumber\\
2f_3^V &=& \Delta g_1^V + \Delta \kappa_V + \lambda_V ~.
\eea
Then Eq.(\ref{alphas}) gives rise to the following constraints 
on
$f_2^{\gamma},~$ $ f_3^{\gamma},$ and $ f_1^{Z}$ :
\bea 
f_1^{\gamma} &=& \eta f_2^{\gamma} ~,
\nonumber\\
f_2^{Z} &=& f_2^{\gamma}~, 
\nonumber\\
 s_w^2 f_3^{\gamma} +  c_w^2 f_3^{Z} &=&
 c_w^2 f_1^{Z}  + \rho f_2^{\gamma}~,
\label{f3}
\eea 
where $\rho = 2\eta c_w^2-\frac{1}{2}$.

We now return to the system of 
Eq.(\ref{sumrule})
and determine what the relations given above 
imply for the 
$\sigma_i$ observables. Feeding the first two relations of 
 Eq.(\ref{f3}) into the system of Eq.(\ref{sumrule}) the latter can now
 be solved  for the two unknown quantities
$f_1^Z$ and $f_2^{\gamma}$ :
\bea
a_1\sigma_1+ \sigma_2+a_4\sigma_4 &=&
 (\eta A_3+A_7+A_8)f_2^{\gamma} + A_4 f_1^Z ~,
\nonumber\\
\sigma_3 &=&  \eta(A_9 - A_{10}) f_2^{\gamma} + A_{10} f_1^Z ~. 
\eea 
Defining
\bea
B_1&=&  \eta A_3 + A_7+A_8 \nonumber\\
B_2 &=&  \eta(A_9 - A_{10}) \nonumber\\ 
\Delta &=& B_1A_{10}- B_2A_4 ~, 
\eea 
we obtain the solution 
\bea
f_2^{\gamma}  & = &
- [a_1 A_{10}\sigma_1+A_{10}\sigma_2
-A_{4}\sigma_3 +a_4A_{10}\sigma_4]\Delta^{-1}~,
\nonumber\\ 
f_1^Z  & = &  -[a_1 B_2 \sigma_1 + B_2 \sigma_2
-B_1 \sigma_3 +a_4 B_2  \sigma_4 ]\Delta^{-1} ~.
\eea
Finally, substituting the above solutions and 
the solutions for $f^{\gamma}_3$, $f^Z_3$ from 
Eq.(\ref{anmagmom}) into the constraint of the third relation of Eq.(\ref{f3}) 
the latter assumes the following form in terms of the $\sigma_i$ 
observables :
\be
h_1 \sigma_1 +h_2 \sigma_2 +h_3 \sigma_3 +h_4 \sigma_4 =0 ~, 
\label{kati}
\ee
where
\bea
h_1  &=& \Delta(s_w^2\gamma_1+c^2 z_1) + a_1(c^2_wB_2-\rho A_{10})~,
\nonumber\\
h_2  &=& c^2_w B_2- \rho A_{10}~, \nonumber\\
h_3  &=& - c^2_w B_1+ \rho A_{4}~,\nonumber\\
h_4  &=& \Delta(s_w^2\gamma_4+c^2 z_4) + a_4(c^2_wB_2-\rho A_{10}) ~.
\eea
Eq.(\ref{kati})
constitutes the prediction of this approach; it can be cast 
 in the alternative form 
\be
 \int_{-1}^{1} dx (z-x)  
\left(\frac{d\sigma^{exp}}{dx}-\frac{d\sigma^{0}}{dx}\right)
H_{3}(x,s) =0~, 
\ee
with 
\be
H_{3}(x,s) = h_1\widetilde{P}_1 + h_2 \widetilde{P}_2 + 
h_3 \widetilde{P}_3+ h_4\widetilde{P}_4
~.
\ee

A particular case of the linear realization of the symmetry is 
the so called HISZ scenario, proposed in \cite{HISZ},
where the third operator 
(usually denoted ${\cal O}$$_{WWW}$) in Eq.(\ref{d6}) is missing.  
Equivalently, one sets $a_W=0$ or $\lambda_V=0$ in Eq.(\ref{CPeff}). 
Then the relations imposed on the 
anomalous couplings become:
\bea
f_1^{\gamma}=f_2^{\gamma}=f_2^{Z}=0 \label{hisz1} \\
 s_w^2 f_3^{\gamma} +  c_w^2 f_3^{Z} =
 c_w^2 f_1^{Z}~, 
 \label{hisz2}
\eea
and Eq.(\ref{sumrule}) transforms to the over-constrained system 
\bea
a_1\sigma_1+\sigma_2+a_4\sigma_4 &=& A_{4} f_1^{Z}~,
\nonumber \\
\sigma_3&=&A_{10} f_1^{Z}~.
\eea
This provides the solution 
\be
f_1^{Z}=\frac{\sigma_3}{A_{10}}
\label{hisz3}
\ee
and the additional constraint 
\be
a_1A_{10}\sigma_1+A_{10}\sigma_2-A_{4}\sigma_3
+a_4A_{10}\sigma_4=0~.
\ee
Evidently in this case where the electric charge 
form factor has been strictly set to 1 while, at the same time
$f_2^Z=0$,
$\sigma_3$ measures directly the weak charge of the $W$. 

The initial constraint of Eq.(\ref{hisz2}) in terms of the $\sigma_i$ reads 
\be
(s_w^2\gamma_1+c_w^2 z_1)\sigma_1-\frac{c_w^2}{A_{10}}\sigma_3
+(s_w^2\gamma_4+c_w^2 z_4)\sigma_4=0~.
\ee

Both constraints can be cast in the equivalent form :
\be
 \int_{-1}^{1} dx (z-x)  
\left(\frac{d\sigma^{exp}}{dx}-\frac{d\sigma^{0}}{dx}\right)
H_{i}(x,s) = 0~, ~~~~~~~~i=4,5
\label{hisz1sr} 
\ee
with
\bea
H_{4}(x,s) & = & a_1A_{10}\widetilde{P}_1+A_{10}\widetilde{P}2-A_{4}\widetilde{P}_3
+a_4A_{10}\widetilde{P}_4~, 
\nonumber \\
H_{5}(x,s)&=& (s_w^2\gamma_1+c_w^2 z_1)\widetilde{P}_1
-\frac{c_w^2}{A_{10}}\widetilde{P}_3
+(s_w^2\gamma_4+c_w^2 z_4)\widetilde{P}_4~.
\eea

In the very heavy Higgs case (or equivalently if the Higgs is absent),
where the symmetry is realized non-linearly,
the Higgs doublet is replaced by a unitary matrix
$U\equiv\mbox{exp} (i{\bf \omega \cdot \tau}/v)$, 
where the  $\omega_i$ are the would-be Goldstone bosons,
and the appropriate
matrix form of the covariant derivative is implied.
It is easy to see that the sum rule obtained in this case
assumes again the simple form of
Eq.(\ref{hisz1sr}). 
Indeed, naive dimensional analysis \cite{Georgi}
 suggests that the $\lambda_V$ couplings are suppressed by additional
powers 
 of the new physics scale ($M_W^4/\Lambda^4$) and are expected to be
negligible
 with respect to $\Delta g_1^V$ and $\Delta\kappa_V$. Thus they are set 
 to zero and there remain again three free parameters $\Delta g_1^Z$, 
 $\Delta\kappa_{\gamma}$ and $\Delta\kappa_Z$. 
  Accordingly, for the $f_i$ 
we obtain the  relation of 
 Eq.(\ref{hisz1}), but not the second relation of Eq.(\ref{hisz2}),
  and the constraint
becomes simply 
 \be
 \int_{-1}^{1} dx (z-x)  
\left(\frac{d\sigma^{exp}}{dx}-\frac{d\sigma^{0}}{dx}\right)
H_{4}(x,s) = 0 ~.
\ee

\setcounter{equation}{0}
\section{Conclusions}

In this paper  we  have presented a set   of  sum rules   relating the
anomalous   gauge   boson couplings to  the   unpolarized differential
cross-section of the  process $e^{+}e^{-} \to  W^{+}W^{-}$.  These sum
rules involve only those anomalous couplings which separately conserve
$C$   and $P$, and have  been  derived under  the  assumption that the
produced  $W$ bosons  are  strictly on-shell.   For this  case we have
defined four  observables, called  $\sigma_i$, $i=1,2,3,4$,  which are
linear combinations of the deviations  of the triliner gauge couplings
from their  SM values.   The  $\sigma_i$ observables can be  extracted
from the  experimentally measured differential   cross section by  (i)
subtracting   out the known   tree-level value of  the  
differential cross  section in the   absence of anomalous couplings;
(ii) multiplying  the   remainder by the  angular   dependence  of the
$t$-channel   propagator; the latter   is also an experimentally known
quantity, since it only depends on the  center-of-mass energy $s$, and
scattering    angle   $\theta$;    (iii)    convoluting the  resulting
expressions   with   four  appropriately  constructed  polynomials in
$\cos\theta$ of maximum degree three.

The  role of these  observables is twofold  : On the  one hand, two of
these observables, namely $\sigma_1$, and $\sigma_4$, represent direct
measurements of  the magnetic moments  $G_M^{\gamma}$ and $G_M^{Z}$ of
the  $W$, while the other two  constitute model independent contraints
(sum rules) between the  remaining anomalous couplings.  Thus, the two
magnetic moments of the $W$ boson  can be {\sl separately } determined
from  the measurement of the   unpolarized differential cross section. 
On the other hand, these observables  are useful for testing dynamical
models which  predict sizeable anomalous  couplings.  This is a direct
consequence of the fact that some of those models predicts constraints
between  the  anomalous couplings,  which,  in  turn, can  be directly
translated  to relations among the   $\sigma_i$ observables.  We  have
demonstrated this possibility in the context of a composite model,
and a model based on an effective gauge-invariant Lagrangian .

Although  we have restricted our  discussion to couplings that respect
$C$ and $P$, this method can be followed step by step also in the case
where the trilinear vertex assumes its  most general form.  Of course,
the system of equations that would correspond to Eq.(\ref{sigma}) will
be modified; in particular, it is not clear whether one would still be
able to isolate $G_{M}^{\gamma}$ and $G_{M}^{Z}$, as happened in the
simpler case we have considered here.
 
It  would  be interesting  to determine  how  the analysis and results
presented here are  modified by the  off-shellness effects of the $W$. 
This next step   may  be  necessary in  view   of the  fact  that  the
cross-section  for on-shell $W$ pair  production  will not be measured
with sufficient accuracy at LEP2.  Such an analysis is complicated not
only due    to the  large   number of  additional   tree-level Feynman
diagramms contributing to the process  $e^{+}e^{-} \to WW \to 4f$, but
also by the fact that the (off-shell) $W$s may be resonant \cite{WG1}.
Calculations in this direction are already in progress.

\end{document}